\documentclass{ptephy_v1}
\usepackage{amsmath,amssymb}
\usepackage{bm}

\preprintnumber{XXXX-XXXX} 

\newcommand\x{\mathbf{x}}
\newcommand\p{\mathbf{p}}
\renewcommand\k{\mathbf{k}}
\newcommand\+{\dagger}

\renewcommand\>{\rangle}
\renewcommand\d{\partial}

\begin{document}

\title{The Dirac composite fermion of the fractional quantum Hall effect}


\author{Dam Thanh Son}
\affil{Kadanoff Center for Theoretical Physics, University of Chicago, Chicago, IL 60637, USA \email{dtson@uchicago.edu}}


\begin{abstract}%
We review the recently proposed Dirac composite fermion theory of the
half-filled Landau level.  This paper is based on a talk given at the
Nambu Symposium at the University of Chicago, March 11--13, 2016.
\end{abstract}

\subjectindex{I96}

\maketitle

\section{Introduction}

The fractional quantum Hall effect (FQHE) was discovered in
1982~\cite{Tsui:1982yy}, only a couple of years following the
discovery of the integer quantum Hall effect (IQHE).  One of the most
nontrivial problems of condensed matter physics, the FQHE has
attracted the attention of theorists ever since.  (One of the earliest
and most influential works is the one by
Laughlin~\cite{Laughlin:1983fy}.)  This paper surveys the most recent
progress in the understanding of one particular, but very important,
aspect of the FQHE: the composite fermion in the half-filled Landau
level~\cite{Halperin:1992mh}.  In particular, we will review the
arguments leading to the Dirac composite fermion
theory~\cite{Son:2015xqa}.

The quantum Hall problem is attractive for theorists partly because of
its very simple starting point: a Hamiltonian describing particles
moving on a two-dimensional plane, in a constant magnetic field, and
interacting with each other through a two-body potential,
\begin{equation}\label{H-TOE}
  H = \sum_{a=1}^N \frac{(\p_a +  \mathbf{A}(\x_a))^2}{2m}
  + \sum_{\<a,b\>} V(|\x_a-\x_b|).
\end{equation}
Here, $\mathbf{A}$ is the gauge potential corresponding to a constant
magnetic field.  The two-body potential $V$ is normally taken to be
the Coulomb potential $V(r)=e^2/r$, but one believes many results are
valid for a large class of repulsive interactions.  The quantum Hall
states are characterized by many physical properties, including
quantized Hall resistivity, vanishing longitudinal resistivity, bulk
energy gap, edge modes, etc.  For the purpose of this article, we take
the existence of an energy gap to be the defining property of the
quantum Hall states.  A very simplified summary of the experimental
situation is as follows: for certain values of the filling factor,
defined as
\begin{equation}
  \nu = \frac \rho{B/2\pi}\,,
\end{equation}
where $\rho$ is the two-dimensional electron density, the system is in
one of the quantum Hall states with an energy gap.  The values of
$\nu$ for which there is a gap are either integers, in which case we
have IQHE, or rational numbers, which correspond to FQHE.

The existence of a gap for integer $\nu$ can be understood on the
basis of the approximation of noninteracting electrons.  In a
magnetic field $B$, the energy eigenvalues of the one-particle
Hamiltonian are organized into Landau levels,
\begin{equation}
 E_n = \frac Bm \left( n+\frac12 \right).
\end{equation}
The degeneracy of each Landau level is $B/2\pi$ per unit area.  At
integer $\nu$, states with $n\le\nu$ are filled and those with $n>\nu$
are left empty.  The system then has a gap equal to the spacing
between Landau levels, which is $\omega_c=B/m$.

In contrast to the IQHE, the fractional quantum Hall effect cannot be
understood from the noninteracting limit.  For example, when
$0<\nu<1$, the lowest Landau level (LLL, $n=0$) is partially filled,
so the noninteracting Hamiltonian has an exponentially large (in the
number of electrons) ground state degeneracy.  The miracle of the FQHE
is that for certain rational values of $\nu$, interactions between
electrons lead to a gap.

There are two energy scales in the FQH problem.  The first scale is
the cyclotron energy $\omega_c=B/m$, while the second scale is the
interaction energy scale.  In the case of the Coulomb interaction, the
latter energy scale can be estimated as the potential energy
between two neighboring electrons,
\begin{equation}
  \Delta = \frac{e^2}r \sim {e^2}{\sqrt B}.
\end{equation}
The FQH problem is usually considered in the limit
$\Delta\ll\omega_c$.  This limit is reached experimentally by taking
$B\to\infty$ at fixed $\nu$; theoretically, it is also reached by
taking $m\to0$ at fixed $B$.  When $\Delta\ll\omega_c$ one can ignore
all Landau levels above the lowest one, and the problem can be
reformulated as pertaining to a Hamiltonian which operates only on the
LLL,
\begin{equation}\label{H-projected}
  H = \mathcal P_{\rm LLL} \sum_{\<a,b\>} V(|\x_a-\x_b|),
\end{equation}
where $P_{\rm LLL}$ is the projection to the lowest Landau level.
This extremely simple Hamiltonian, believed to underlie all the
richness of FQH physics, cannot be solved by traditional methods of
perturbation theory due to the lack of a small parameter.  In
particular, there is only one energy scale---the Coulomb energy scale
$\Delta$.  The FQH problem is essentially nonperturbative.

\section{Flux attachment}

One of the most productive ideas in FQH physics has been the idea
of the composite fermion (CF).  The notion of the CF itself is based
on another concept called flux attachment~\cite{Arovas:1985yb}, which
was applied to the FQHE in a number of groundbreaking
works~\cite{Zhang:1988wy,Jain:1989tx,Fradkin:1991wy,Halperin:1992mh}.
I will now review the standard textbook field theory of the composite
fermion, although later on I will argue that it needs some nontrivial
modification to become the correct low-energy effective theory.

In the FQH case, one ``attaches'' an even number (in the simplest
case, two) of magnetic flux quanta to an electron, transforming it to
a new object called the ``composite fermion.''  In field theory
language, one starts from a theory of interacting electrons $\psi_e$
in (2+1) dimensions in a background magnetic field
\begin{equation}\label{L-orig}
  \mathcal L = i \psi_e^\+ (\d_t - i A_0) \psi_e
   - \frac1{2m}|(\d_i-iA_i)\psi_e|^2 + \cdots
\end{equation}
where $\cdots$ stands for interaction terms, and ``derives,'' following
a certain formal procedure, a new Lagrangian for the composite fermion
$\psi$,
\begin{equation}\label{L-cf}
  \mathcal L = i\psi^\+ (\d_t -iA_0 + ia_0)\psi
  - \frac1{2m} |(\d_i - iA_i + ia_i)\psi|^2 +
  \frac12 \frac1{4\pi} \epsilon^{\mu\nu\lambda}
  + a_\mu \d_\nu a_\lambda \cdots
\end{equation}
The Chern--Simons term in Eq.~(\ref{L-cf}) encodes the idea of flux
attachment.  In fact, the equation of motion obtained by
differentiating the action with respect to $a_0$ reads
\begin{equation}\label{flux_att}
  2 \psi^\+ \psi = \frac b{2\pi}\,,
  \qquad b=\bm{\nabla}\times \mathbf{a},
\end{equation}
which means that the magnetic fluxes of the dynamic gauge field
$a_\mu$ are tied to the location of the composite fermions, with two
units of fluxes per particle.

There are two features of the field theory~(\ref{L-cf})---which will
be called the HLR field theory after Halperin, Lee, and Read who used
it to study the half-filled Landau level~\cite{Halperin:1992mh}---which
are rather trivial but worth listing here for future reference:
\begin{itemize}
\item The number of composite fermion is the same as the number of
  electrons.  It cannot be otherwise if the composite fermion results
  from attaching magnetic fluxes to an electron.
\item The action contains a Chern--Simons term for $a_\mu$.  As
  demonstrated above, this term encodes in mathematical terms the idea
  of flux attachment.
\end{itemize}

In the literature, it is often stressed that transformation from
(\ref{L-orig}) to (\ref{L-cf}) can be done in an exact way (see, e.g.,
Ref.~\cite{Fradkin:1991wy}).  ``Conservation of difficulty'' then
implies that the theory~(\ref{L-cf}) cannot be solved exactly.  To
make any progress at all, one has to start with some approximation
scheme, and in every work so far this has been the mean field
approximation where one replaces the dynamical gauge field $a_\mu$ by
its average value determined from Eq.~(\ref{flux_att}).  Since in the
Lagrangian (\ref{L-cf}) the gauge fields $A$ and $a$ enter through the
difference $A-a$, and the density of the composite fermions is the
same as the density of the original electrons, the effective average
magnetic field acting on $\psi$ is
\begin{equation}
  B_{\rm eff} = B - \<b\> = B-4\pi\rho.
\end{equation}
Translated to the language of the filling factors, 
\begin{equation}
  \nu = \frac\rho{B/2\pi}\,, \qquad \nu_{\rm CF} = \frac\rho{B_{\rm eff}/2\pi}\,,
\end{equation}
the equation becomes
\begin{equation}
  \nu_{\rm CF}^{-1} = \nu^{-1}-2.
\end{equation}
In particular, the values $\nu=\frac{n}{2n+1}$ map to $\nu_{\rm
  CF}=n$.  In this way we have mapped the FQH problem for the electron
to the IQH problem for the composite fermions, which gives an
``explanation'' for the emergence of an energy gap.  Experimentally,
one finds quite robust quantum Hall plateaux at these values of $\nu$,
up to $n\approx10$.

Another sequence of quantum Hall plateaux are found at
$\nu=\frac{n+1}{2n+1}$.  Now $\nu>\frac12$ so the effective average
magnetic field $B_{\rm eff}$ is negative, i.e., points in the
direction opposite to the direction of the original $B$.  The
composite fermion still forms IQH states, with $n+1$ filled Landau
levels ($\nu_{\rm CF}=-(n+1)$).  Together, the two series of FQH
plateaux at $\nu=\frac n{2n+1}$ and $\nu=\frac{n+1}{2n+1}$ are called
the Jain sequences of plateaux.

One of the most spectacular successes of the composite fermion theory
is the prediction of the nature of the $\nu=\frac12$ state (the
half-filled Landau level)~\cite{Halperin:1992mh}.  At this filling
fraction, the average effective magnetic field is equal to 0, and the
composite fermion should form a gapless Fermi surface.  HLR theory
thus predicts that the low-energy excitation is the fermionic
quasiparticle near the Fermi surface.  There is strong experimental
evidence that this is indeed the
case~\cite{Willett:1990,Kang:1993,Goldman:1994zz}.  These experiments
give the strongest evidence that the composite fermion is a real
physical object---a quasiparticle near half filling---and not just a
mathematical construct.

Despite its astounding success, the quantum field theory~(\ref{L-cf})
has been criticized on various grounds.  The criticism leveled most
often against the theory~(\ref{L-cf}) is the lack of any information
about the projection to the lowest Landau level.  In particular, the
energy gap predicted by the mean-field picture is $B_{\rm eff}/m$,
which for generic $\nu$ is of order $\omega_c$, but not $\Delta$.  To
remedy the issue, one has to assume that the energy gap is determined
by an effective mass $m_*$, postulated to be parametrically
$B/\Delta$.  In particular, $m_*$ is assumed to remain finite in the
limit $m\to0$.

In my view, there are in reality two energy scale problems.  The first
problem, which I would call the ``grand problem'' of energy scale, is
to derive, from microscopic calculations, the finite value of $m_*$ in
the limit $m\to0$.  The second problem, the more modest ``little
problem'' of energy scale, is to make the low-energy effective field
theory with $m_*$ consistent with the fundamental symmetries of the
original theory of electrons with a much smaller mass.

The ``grand problem'' is the one that attracts most attention.  We
note here a few past attempts to address
it~\cite{Shankar:1997zz,Pasquier:1998sre,Read:1998dn}.  However
important it is, it will not concern us if our ambition is limited to
capturing the low-energy phenomenology, i.e., the physics at energy
scales much smaller than $\Delta$.  The effective mass $m_*$ would
appear simply as an input parameter in a low-energy effective field
theory, and we will simply postulate that such an effective mass
arises somehow as a result of the renormalization group flow from a
UV scale above $\omega_c$ to an IR scale below $\Delta$.
The ``little problem'' of energy scale is a fully low-energy question,
and it can now be solved, in principle, by using the
Newton--Cartan formalism (see, e.g.,
Refs.~\cite{Son:2013rqa,Jensen:2014aia,Geracie:2015xfa,Geracie:2015dea}).

However, the most recent progress in the physics of the half-filled
Landau level has arrived from an attempt to address another problem,
usually regarded as less important and subordinate to the energy scale
problem: the lack of particle--hole (PH) symmetry.

\section{The problem of particle--hole symmetry}

A system of nonrelativistic particles interacting through a two-body
interaction has two discrete symmetries: parity, or spatial reflection
($x\to x$, $y\to -y$), which we denote as $P$, and time reversal,
which will be called $T$.  In a constant uniform magnetic field both
$P$ and $T$ are broken, but $PT$ is preserved.  But in the lowest
Landau level limit ($\Delta\ll\omega_c$), the projected
Hamiltonian~(\ref{H-projected}) has an additional discrete symmetry:
the particle--hole symmetry, first considered in
Ref.~\cite{Girvin:1984zz}.

To define the particle--hole symmetry, one chooses a particular basis of
LLL one-particle states $\psi_k(x)$.  This basis defines the electron
creation and annihilation operators $c_k^\+$, $c_k$.  The many-body
LLL Fock space is obtained by acting products of creation operators on
the empty Landau level $|\textrm{empty}\>$.

Particle--hole conjugation, $\Theta$, is defined as an antilinear
operator, which maps an empty Landau level to a full one:
\begin{equation}
  \Theta: |\textrm{empty}\> \to |\textrm{full}\> = \prod_{k=1}^M
  c_k^\+ |\textrm{empty}\>,
\end{equation}
where $M$ is the number of orbitals on the LLL.  It also maps a
creation operator to an annihilation operator, and vice versa:
\begin{equation}
  \Theta: c_k^\+ \leftrightarrow c_k .
\end{equation}

One can show that the projected Hamiltonian maps to itself, up to the
addition of a chemical potential term,
\begin{equation}
  \Theta: H_{\rm LLL} \to H_{\rm LLL} - \mu_0 \sum_k c^\+_k c_k ,
\end{equation}
where $\mu_0$ depends on the interaction $V$.  This means that for
$\mu=\mu_0/2$, the Hamiltonian $H_{\rm LLL}-\mu N$ maps to itself: at
this chemical potential the Hamiltonian is particle--hole symmetric.

Under particle--hole conjugation the filling factor $\nu$ transforms
as
\begin{equation}
  \nu \to 1-\nu .
\end{equation}
In particular $\nu=1/2$ maps to itself under PH conjugation: the
half-filled Landau level is at the same time half empty.  Moreover,
$\nu=\frac n{2n+1}$ maps to $\nu=\frac{n+1}{2n+1}$: the two Jain
sequences of quantum Hall plateaux form pairs that map to each other
under PH conjugation: $\nu=1/3$ and $\nu=2/3$, $\nu=2/5$ and
$\nu=3/5$, etc.

Let us now ask what the discrete symmetries of the HLR field
theory~(\ref{L-cf}) are.  It is easy to see that there is only one
such symmetry, $PT$.  The Chern-Simons theory does not have any
discrete symmetry that can be associated with particle--hole
conjugation.  This reflects on the asymmetry in the treatment of
quantum Hall plateaux: the $\nu=\frac n{2n+1}$ is described by an
integer quantum Hall state where the CFs fill $n$ Landau levels, while
its PH conjugate $\nu=\frac{n+1}{2n+1}$ by $n+1$ filled Landau levels.

The Fermi liquid state with $n=1/2$ presents a particularly baffling
problem for particle--hole symmetry.  Naively, one expects PH
conjugation to map a filled state to an empty state an vice versa.
This would mean that the Fermi disk of the CFs, describing the Fermi
liquid state, maps to a hollow disk in momentum states: the states
with momentum $|\k|>k_F$ are filled, and those with $|\k|<k_F$ are
empty.  This is obviously silly.

The lack of particle--hole symmetry has been recognized as a problem
of the HLR theory from early on.  One aspect of this problem was
noticed in 1997 by Kivelson, Lee, Krotov, and
Gan~\cite{Kivelson:1997}.  When disorders are statistically
particle--hole symmetric, particle--hole symmetry implies that at half
filling $\sigma_{xy}$ is exactly $\frac12(e^2/h)$, but the HLR theory,
in the random phase approximation, implies that $\rho_{xy}=2(h/e^2)$.
These two results disagree with each other when the longitudinal
conductivity $\sigma_{xx}$ (or equivalently, the longitudinal
resistivity $\rho_{xx}$) is nonzero.  From time to time, the issue of
particle--hole symmetry has been brought up in the literature (for
example, it was crucial for the discovery of the anti-Pfaffian
state~\cite{Levin:2007,SSLee:2007}), but no conclusive resolution of
the problem of the lack of PH symmetry in the HLR theory has been
found.

What makes the PH symmetry problem seem hard is that PH symmetry is
not the symmetry of nonrelativistic electrons in a magnetic field [the
  theory~(\ref{H-TOE})].  It only emerges as the symmetry after taking
the lowest Landau level limit [theory~(\ref{H-projected})].  The
particle--hole symmetry of the LLL is not realized as a local operation
acting on fields.

It was commonly thought that the PH symmetry problem is part of the
energy scale problem: PH symmetry becomes exact in the LLL limit,
where the energy scale problem is sharpest.  But in fact, the PH
symmetry problem is easier than the ``grand problem'' of energy scale:
PH symmetry is a question about the low-energy effective field theory,
while the CF effective mass, the object of concern of the energy scale
problem, comes mostly from energy scales above $\Delta$.

One can envision three possible scenarios for the problem of
particle--hole asymmetry of the HLR theory to resolve itself:
\begin{itemize}
\item[(i)] Despite the lack of an explicit PH symmetry, the HLR theory
  has a hidden PH symmetry.
\item[(ii)] Particle--hole symmetry is spontaneously broken, and the
  HLR theory describes only the low-energy excitations around one of
  the two ground states.
\item[(iii)] The effective field theory describing the low-energy
  excitations is different from HLR.  In this theory, particle--hole
  symmetry is explicitly realized.
\end{itemize}
Option (i) cannot be ruled out, but a careful diagrammatic analysis by
Kivelson et al.~\cite{Kivelson:1997} does not seem to reveal any
mechanism under which particle--hole symmetry may be hidden.  How this
can be reconciled with the supposed exactness of the flux attachment
procedure is not clear, but one should remember that the HLR theory,
as applied in practice, makes an additional assumption of the mean
field Fermi liquid as the starting point.  One thing is clear: if one
takes the HLR Lagrangian and declares it (after making some standard
modifications like changing the electron mass $m$ to the effective
mass $m_*$, adding Landau's interactions, etc.) to be the Lagrangian
of a low-energy effective field theory (with a cutoff much smaller
than the Fermi energy), then this effective field theory would show no
indication of particle--hole symmetry.

Option (ii) is self-consistent and was investigated by Barkeshli et
al.~\cite{Barkeshli:2015afa}.  If that is the case, there are two
states at $\nu=1/2$: one corresponds to a Fermi surface of ``composite
particles'' and the other to that of ``composite holes.''  However,
there is no numerical or experimental evidence for this kind of
spontaneous particle--hole symmetry breaking.  In fact, the
experimental result of Ref.~\cite{Baldwin:2014} seems to indicate, at
least naively, that the $\nu=1/2$ Fermi liquid is equally well
interpreted as being made out of ``composite particles'' or
``composite holes.''  There is now strong numerical evidence that the
$\nu=1/2$ state is particle--hole symmetric~\cite{Geraedts:2015pva}.

We will now try to make sense of option (iii).

\section{Dirac composite fermion}

There exists an alternative theory that satisfies particle--hole
symmetry but also preserves all successful phenomenological
predictions of the HLR theory.  This theory is the Dirac composite
fermion theory, first proposed in Ref.~\cite{Son:2015xqa} as the
low-energy effective field theory of the half-filled Landau level.
The essence of the theory is that the composite fermion does not
transform into a ``composite hole'' under particle--hole symmetry, but
remains a composite particle.  Only the momentum of the composite
fermion flips sign under particle--hole conjugation,
\begin{equation}\label{PH_CF}
  \Theta: \k \to -\k.
\end{equation}
Implicitly, we assume that the Fermi disk of the composite fermion
transforms into itself (a filled disk, not a hollow disk).

Equation~(\ref{PH_CF}) is how time reversal usually works.  In the
theory of the Dirac composite fermion, the CF is described by a
two-component spinor field $\psi$, which transforms under PH
conjugation following the formula usually associated with time
reversal,
\begin{equation}
 \psi \to i\sigma_2\psi .
\end{equation}

There are several arguments one can put forward to argue that the
composite fermion has to be a massless Dirac particle.  One argument,
or rather a hint, comes from the CF interpretation of the
Jain-sequence states.  Recall that one problem with the standard CF
picture is that $\nu=\frac n{2n+1}$ corresponds to the composite
fermion filling factor $\nu_{\rm CF}=n$, while $\nu=\frac{n+1}{2n+1}$
maps to $\nu_{\rm CF}=n+1$ (ignoring the sign).  On the other hand,
these two states are PH-conjugate pairs and should be described by the
same filling factor of the composite fermion in any PH-symmetric
theory.  The most naive way to reconcile these different pictures is
to replace the filling factors $\nu_{\rm CF}=n$ and $\nu_{\rm CF}=n+1$
with the average value $\nu_{\rm CF}=n+\frac12$.  But now we have a
problem: we want to map the FQHE in the Jain sequences to the IQHE of
the composite fermions, but is it possible to have an IQH state with
half-integer filling factor?  Indeed it is, if the composite fermion
is a massless Dirac fermion.  Half-integer quantization of the Hall
conductivity is a characteristic feature of the Dirac fermion,
confirmed in experiments with
graphene~\cite{Novoselov:2005kj,Zhang:2005zz}.

The second argument in favor of the Dirac nature of the CF relies on a
property of the square of the particle--hole conjugation operator
$\Theta^2$~\cite{Geraedts:2015pva}.\footnote{Also, M.~Levin and
  D.~T.~Son, unpublished (2015).}  It is intuitively clear that
applying particle--hole conjugation twice maps a given state to
itself, but there is a nontrivial factor of $\pm1$ that one gains by
doing so.

Consider a generic state on the LLL with $N_e$ electrons,
\begin{equation}
  |\psi\> = \prod_{i=1}^{N_e} c_{k_i}^\+ \, |\textrm{empty}\>.
\end{equation}
Then under PH conjugation,
\begin{equation}
  \Theta: |\psi\> \to \prod_{i=1}^{N_e} c_{k_i}^{\phantom{\dagger}} |\textrm{full}\>
  = \prod_{i=1}^{N_e} c_{k_i}^{\phantom{\dagger}} 
    \prod_{j=1}^{M} c_j^\+ |\textrm{empty}\>.
\end{equation}
Applying $\Theta$ again one finds
\begin{equation}
  \Theta^2: |\psi\> \to \prod_{i=1}^{N_e} c_{k_i}^\+ \prod_{j=1}^M c_j
  |\textrm{full}\> =
   \prod_{i=1}^{N_e} c_{k_i}^\+ \prod_{j=1}^M c_j^{\phantom{\dagger}}
  \prod_{k=1}^M c_k^\+ |\textrm{empty}\> = (-1)^{M(M-1)/2}|\psi\>.
\end{equation}

This relationship is quite easy to interpret when $M$ is an even
number: $M=2N_{\rm CF}$.  Then
\begin{equation}\label{Theta2NCF}
  \Theta^2: |\psi\> \to (-1)^{N_{\rm CF}} |\psi\>.
\end{equation}
This formula suggests the following interpretation: $N_{\rm CF}$ is
the number of composite fermions of the state $|\psi\>$, and each
composite fermion is associated with a factor of $-1$ under
$\Theta^2$.  This $-1$ factor is natural for the Dirac fermion.

In order to have a correct $\Theta^2$, we have to identify the number
of composite fermions with half the number of orbitals on the LLL:
$N_{\rm CF}=M/2$, which is \emph{independent} of the number of
electrons $N_e$.  This contradicts the intuitive picture of flux
attachment, in which the composite fermion is obtained by attaching
two units of flux quanta to an electron.  On the other hand, that is
expected: in a theory that treats particles and holes in a symmetric
way, the number of composite fermions has to be in general different
from the number of electrons, otherwise it would have to be equal to
the number of holes as well.

The tentative theory of the composite fermion can be written as follows
\begin{equation}\label{L-dual}
  \mathcal L = i\bar\psi \gamma^\mu(\d_\mu + 2ia_\mu)\psi + \frac1{2\pi}
  \epsilon^{\mu\nu\lambda}A_\mu\d_\nu a_\lambda
  + \frac1{8\pi} \epsilon^{\mu\nu\lambda}A_\mu \d_\nu A_\lambda.
\end{equation}
(with a speed of light which is determined by microscopic physics).
There are two differences between (\ref{L-dual}) and (\ref{L-cf}).
One is the Dirac nature of the composite fermion $\psi$.  The other is
the absence of the Chern--Simons term $ada$ in the Lagrangian: such a
term (as also the mass term for $\psi$), if present, would disallow
any discrete symmetry that could be identified with particle--hole
symmetry.  Interestingly, each such modification to the HLR theory
would shift the filling factors of the Jain-sequence plateaux, but
together the shifts cancel each other and the Jain sequences remain
unchanged, as shown below.

How should one visualize the composite fermion?  In
Ref.~\cite{Son:2015xqa} it was suggested that the CF is better
interpreted as a type of fermionic vortex, arising from a fermionic
particle--vortex duality.  Particle--vortex duality is well known for
bosons~\cite{Peskin:1977kp,Dasgupta:1981zz}, but we are dealing here
with a new duality for fermions.  The salient feature of
particle--vortex duality is that it switches the roles of particle
number and magnetic field.  Differentiating~(\ref{L-dual}) with
respect to $A_0$, one obtains the electron density
\begin{equation}\label{rhob}
  \rho = \frac{\delta S}{\delta A_0} = \frac b{2\pi} +\frac B{4\pi} \,.
\end{equation}
On the other hand, the equation of motion obtained by differentiating
the action with respect to $a_0$ is
\begin{equation}\label{rhoCFB}
   \bar\psi \gamma^0 \psi = \frac B{4\pi}\,,
\end{equation}
i.e., the CF density is set by the external magnetic field.

If one defines the filling factors of the electron and the composite
fermion as
\begin{equation}
  \nu = \frac{2\pi\rho} B\,,\qquad \nu_{\rm CF} = \frac{2\pi\rho_{\rm CF}}b\,,
\end{equation}
then from Eqs.~(\ref{rhob}) and (\ref{rhoCFB}) we find that they are
related by
\begin{equation}
  \nu_{\rm CF} = - \frac1{4(\nu-\frac12)}\,.
\end{equation}
In particular, $\nu=\frac n{2n+1}$ maps to $\nu_{\rm CF}=n+\frac12$,
which is the filling factor of an integer quantum Hall state of the
Dirac fermion.

It should be emphasized that the Dirac nature of the CF does not mean
that there is a Dirac cone for the CF.  The tip of the cone is at
$\k=0$ while the CF, as a low-energy mode, exists only near the Fermi
surface.  The Dirac nature of the CF, strictly speaking, only means
that the fermionic quasiparticle has a Berry phase of $\pi$ around the
Fermi surface.  It is easy to show that such a Berry phase follows
from Eqs.~(\ref{PH_CF}) and (\ref{Theta2NCF}).  The quasiparticle
Berry phase has been identified as an important ingredient of Fermi
liquids~\cite{Haldane:2004zz}, but the possibility of such a phase for
the composite fermion in FQHE has been overlooked in the literature
until very recently.

\section{Consequences of Dirac composite fermion}

The Dirac composite fermion theory has distinct consequences, in
principle verifiable in experiments and numerical simulations.

It is numerical simulations~\cite{Geraedts:2015pva} that provide the
currently most nontrivial test of the Dirac nature of the composite
fermion.  The numerical finding is the disappearance, attributable to
particle--hole symmetry, of the leading $2k_F$ singularity in certain
correlation functions.  
It is well known that for (2+1)D massless Dirac fermion, two-point
correlation functions of time-reversal-invariant operators are free
from the leading $2k_F$ singularity in a generic two-point correlator,
a fact that originates from the quasiparticle Berry phase $\pi$ around
the Fermi surface.  In the half-filled Landau level, the role of time
reversal is played by particle--hole symmetry, therefore to test the
Berry phase one should look for the absence of the leading $2k_F$
singularity in correlation functions of PH symmetric operator.  The
electron density operator $\rho=\psi_e^\dagger\psi_e$ is not PH
symmetric (the deviation of the density from the mean density,
$\delta\rho=\rho-\rho_0$ flips sign under PH conjugation) but one can
easily write down more complicated operators that are PH symmetric,
for example $\delta\rho \nabla^2 \rho$.  In
Ref.~\cite{Geraedts:2015pva} the leading $2k_F$ singularity in the
correlation function of such an operator was shown to disappear when
PH symmetry is made exact (and to reappear when PH symmetry is
violated), confirming the Dirac nature of the composite fermion.

There are also predictions about transport that are, strictly
speaking, consequences of particle--hole symmetry.  If one introduces
the conductivities $\sigma_{xx}$, $\sigma_{xy}$, and the
thermoelectric coefficients $\alpha_{xx}$ and $\alpha_{xy}$,
\begin{equation}
  \mathbf{j} = \sigma_{xx} \mathbf{E} + \sigma_{xy}\mathbf{E}\times
  \mathbf{\hat{z}} + \alpha_{xx} \bm{\nabla} T
  + \alpha_{xy} \bm{\nabla} T \times \mathbf{\hat{z}},
\end{equation}
then, at exact half filling, particle--hole symmetry
implies~\cite{Son:2015xqa,Potter:2015cdn}
\begin{equation}
  \sigma_{xy} = \frac12 \frac{e^2}h\,,\qquad \alpha_{xx}=0.
\end{equation}
A manifestly particle--hole symmetric theory like the Dirac composite
fermion theory reproduces these results automatically.  On the other
hand, the HLR theory, supplemented by the usual approximations to make
it suitable for computation (e.g., the random phase approximation)
would, in general, break both
relationships~\cite{Kivelson:1997,Potter:2015cdn}.

\section{Conclusion}

We have presented arguments in favor of the Dirac nature of the
composite fermion.  The Dirac composite fermion provides a very simple
solution to a number of puzzles that have been plaguing the quantum
field theory of the composite fermion for a long time.

A simple demonstration that the Dirac composite fermion emerges from
the dynamics of interacting electrons on the lowest Landau level is
still lacking. (For a recent attempt to address this question see
Ref.~\cite{Murthy:2016jnc}.) One may wonder how the flux attachment
procedure, supposed to be exact, can lead us to something so different
from Eq.~(\ref{L-cf}).  The situation becomes less puzzling if one
remembers that the Lagrangian (\ref{L-dual}) is a low-energy effective
Lagrangian, while the action of the type~(\ref{L-cf}) obtained from
the exact flux attachment procedure contains information about all
energy scales.  One may also be bothered by the emergence of a Dirac
fermion out of the initial nonrelativistic fermion.  Here again, the
situation is not as strange as it sounds: what is important is not
really the nonrelativistic Hamiltonian~(\ref{H-TOE}), but the LLL
projected Hamiltonian~(\ref{H-projected}), which applies equally well
if the original fermion is a Dirac fermion (e.g., the gapless mode on
the surface of a topological insulator).  In this case the duality is
one between two theories, both involving Dirac fermions.

Going beyond quantum Hall physics, a very interesting possibility is
that the duality between the free Dirac fermion (the electron theory)
and Dirac fermion interacting with a gauge field is valid even at zero
magnetic field.  Such a duality would have consequences for
interacting surfaces of topological insulators: for example, the
so-called T-Pfaffian
state~\cite{Wang:2013uky,Bonderson:2013pla,Chen:2013jha,Metlitski:2015bpa},
otherwise difficult to derive, could be understood simply from the
dual picture (the quantum Hall analog of this state is the state
called PH-Pfaffian in Ref.~\cite{Son:2015xqa} and involves BCS pairing
of Dirac composite fermions in the $s$-wave channel).  Much effort has
been made to derive such
duality~\cite{Metlitski:2015eka,WangSenthil1,WangSenthil2,Mross:2015idy,Karch:2016sxi,Seiberg:2016gmd}.
In one approach, one discretizes the system in one spatial dimension
and utilizes (1+1)D bosonization~\cite{Mross:2015idy}.  In another
approach, the duality between the two fermion theories appears as one
particular case of a whole web of (2+1)D dualities which can be
derived from an elemental duality between a bosonic field theory and a
fermionic field theory~\cite{Karch:2016sxi,Seiberg:2016gmd},
establishing a connection with an extensive literature on duality
between (2+1)D Chern--Simons theories (see, e.g.,
\cite{Aharony:2015mjs}).  The latter approach, in particular,
clarifies issues related to the parity anomaly matching.  It is
unclear, however, if a single two-component fermion coupled to a
dynamical gauge field is stable with respect to spontaneous symmetry
breaking.  Numerical efforts are required to settle this question.
There is a claim that QED$_3$ does not spontaneously generate a gap
for two flavors of two-component fermion~\cite{Karthik:2015sgq}, in
contrast to the general belief.  The situation with one flavor is not
clear.

According to P.~Freund~\cite{Freund:2015nts}, Nambu was fascinated
with the philosophy of science of Mitsuo Taketani, according to which
scientific development passes through three stages: Phenomenon,
Substance, and Essence.  In the story that we have just surveyed, I
guess Nambu would pick the FQH plateaux as the Phenomenon and the
composite fermion as the Substance.  Are we catching, in the fermionic
particle--vortex duality and other field--theoretic dualities in
(2+1)D, a glimpse of the Essence?
\medskip

\noindent
\textbf{Acknowledgments}
\smallskip

\noindent
This work is supported, in part, by U.S.\ DOE
grant No.\ DE-FG02-13ER41958, ARO MURI grant No.\ 63834-PH-MUR, and a
Simons Investigator Grant from the Simons Foundation.  Additional
support was provided by the Chicago MRSEC, which is funded by NSF
through grant DMR-1420709.

\end{document}